\documentclass[plb,draft,preprint,showpacs,groupedaddress]{revtex4-1}

\usepackage{amsmath}
\usepackage{amsfonts}
\usepackage{amssymb}
\usepackage{geometry}
\usepackage[english]{babel}

\begin{document}
\bibliographystyle{apsrev4-1}
\message{}

  \title{A model of eternal accelerated expansion without particle horizon}

  \author{Zi-Liang Wang} 
  \author{Jian-Bo Deng} \email[Jian-Bo Deng: ]{dengjb@lzu.edu.cn} 

  \affiliation{Institute of Theoretical Physics, LanZhou University,
    Lanzhou 730000, P. R. China}

  \date{\today}

  \begin{abstract}
  In our previous paper \cite{8}, we proposed a cosmological model from the emergence of space, which possesses a significant character of evaluating the vacuum energy from the Hubble constant and the age of universe. And one problem of this model is that there is no inflation in the early universe. In this paper, we aim at resolving this problem which leads us to a rather surprising conclusion that our cosmological model can avoid the horizon and flatness problems.
  \end{abstract}


  \maketitle

  \section{Introduction}

The standard model of universe requires initial conditions which are very implausible for two reasons \cite{3}: (1) The early universe seems to be highly homogeneous, in spite of the fact that separated regions were causally disconnected(this is called horizon problem); (2) the initial curvature of the universe must be extremely small to produce a flat universe today (this is named flatness problem). The original inflation was proposed by Guth in 1980 to solve these problems \cite{3}. The third classic cosmological problem is called monopoles problem. However, the most serious of the above three problems is the horizon problem and there are possible solutions of the flatness and monopole problems that do not rely on inflation \cite{2}.



In our last paper \cite{8}, inspired by Padmanabhan's idea that the spatial expansion of universe is due to the difference between the surface degrees of freedom (DOF) and the bulk DOF in the region of emerged space (For more investigations about this idea, see Refs.~\cite{cai2012emergence,yang2012emergence,ai1,sheykhi2013friedmann}),
we proposed a cosmological model from the emergence of space, which is asymptotically de Sitter and possesses a significant character of evaluating the vacuum energy from the Hubble constant and the age of universe. Nonetheless, there are two problems for the model. The first problem that we have mentioned in the last paper is that there is no inflation in the early universe.  
And the second is that there is no deceleration phase which means the expansion of the universe is always accelerating. In this paper, we aim at resolving the first problem. And this effort surprisingly brings the avoidance of the horizon and flatness problems to our cosmological model.

This paper is organized as following. The basic equations of the standard model and the horizon and flatness problems are summarized In Sec.~II, together with a brief introduction about inflationary theory attached. In Sec.~III, after presenting a brief review of our previous work, we demonstrate how our model can eliminate the horizon and flatness problems. Sec.~IV is reserved for conclusions and discussions.

\section{the standard model of universe}

In the standard model, the universe is assumed to be isotropic and homogeneous, therefore can be described by Robertson-Walker metric~\cite{1,2,3}:
\begin{equation}
\label{1} ds^2=-dt^2+a^2(t)\left[ \frac{dr^2}{1-kr^2}+r^2(d \theta ^2 +sin^2\theta d \varphi ^2)\right],
\end{equation}
where $k=+1$ for the 3-sphere, $k=0$ for flat space, and $k=-1$ for the hyperboloid. So, the evolution of scale factor $a(t)$ is governed by Friedmann equations:
\begin{equation}
\label{2}  3H^2+\frac{3k}{a^2}=8\pi  \rho
\end{equation}
\begin{equation}
\label{3} 3\ddot{a}=-4\pi  a (\rho + 3p)
\end{equation}
where $H\equiv \dot{a}/a$ is the Hubble constant. 

Now we will explain the puzzles in the standard model, namely horizon and flatness problems.
For explaining the horizon problem, we need the concept of particle horizons. Particle horizons limit the distance at which past events can be observed \cite{4},  the proper distance of the horizon is given by
\begin{equation}
\label{4} d_{max}(t)=a(t)\int_0 ^t \frac{dt'}{a(t')}\qquad.
\end{equation}
Thus there is a particle horizon unless the integral $\int dt/a(t)$ does not converge at $t=0$. It does converge in conventional cosmological theories \cite{2}. For instance, during the radiation-dominated era $a(t)\propto t^{1/2}$, so $d_{max}(t)=2t$. 
From the cosmic microwave background, we believe that the present universe is homogeneous and isotropic to a very high degree of precision. 
Considering two microwave antennas pointed in opposite directions, which are receiving radiation at the time of hydrogen recombination ($t_{\gamma d} \approx10^5$ years), one can almost get the information without difference. However, at that time of emission, these two sources were separated from each other by over 90 horizon lengths \cite{5}. In other words,
\begin{equation}
\frac{d_{max}(t_{\gamma d})}{D_{12}(t_{\gamma d})}\approx \frac{1}{90}\ \ ,
\end{equation}
where $D_{12}(t_{\gamma d})$ is the proper distance between the two antennas at $t_{\gamma d}$. It is hard to understand how two regions over 90 horizon lengths apart came to be at the same temperature at the same time. This is the horizon problem.

The second puzzle is flatness problem which was first mentioned by Dicke and Peebles \cite{6}. For any value of the Hubble constant $H$, we can define a critical density 
\begin{equation}
\label{5} \rho_{crit}\equiv \frac{3H ^2}{8\pi } 
\end{equation}
According to Eq.~\eqref{2}, whatever we assume about the constituents of the universe, the $k$ will be $+1$ or $0$ or $-1$ according to whether the present density $\rho_0$ is greater than, equal to, or less than the present $\rho_{0,crit}$  \cite{2}. The ratio is called $\Omega_0$ ($\Omega \equiv \rho/\rho_{crit}$). To compare $\Omega$ with 1, we define $\epsilon \equiv |1-\Omega^{-1}|$ and we can use Eq.~\eqref{2} and have 
\begin{equation}
\label{6} \epsilon(t)=\left|\frac{\rho -\rho_{crit}}{\rho}\right|=\frac{3|k|}{8\pi \rho(t)a^2(t)}\ \ .
\end{equation}    
Unless a flat universe ($k=0$), the $\epsilon$ will increase in the universe dominated by matter ($\rho \propto a^{-3}$) or radiation ($\rho  \propto a^{-4}$), 
\begin{equation}
\label{s1}
\epsilon \propto(\rho a^2)^{-1}= \begin{cases}
(\rho a^4)^{-1}a^2 \propto a^2,\ \ (radiation)\\

(\rho a^3)^{-1}a \propto a,\ \ (matter)
\end{cases}\ \ .
\end{equation}
However, the experimental data from the Wilkinson Microwave Anisotropy Probe (WMAP) constrain $\Omega_0$ to be 1 within 1$\%$ \cite{7}. In other words the $\epsilon$ is currently less than 0.01 which means the universe must have been very near to $\epsilon=0$ in the past \cite{5}. This is the flatness problem.

Since these problems can be solved by inflation theory, we will make a brief introduction about it. 
In most inflationary models the energy density $\rho$ is approximately constant(hence $\rho =-p$), leading to exponential expansion of the scale factor \cite{10}
\begin{equation}
\label{12} a(t)\propto e^{\chi t}
\end{equation}
where $\chi=\sqrt{\frac{8\pi}{3}G\rho}\approx 10^{34}$ s$^{-1}$. The inflation approximately begins at $10^{-34}s$ and lasts for about $10^{-32}s$. So during inflation, the scalar factor has increased by $e^{100}$ times, and the $d_{max}(t)$ also has a huge increase. In this case, for a given $D(t_0)$ ($D(t_0)$ means the present diameter of the observed universe  and $D(t) \propto a(t)$ is the value of $D(t_0)$ at $t$), the inflation can give a value of $D(t)$ which is $e^{100}$ times smaller than the value given by the standard model when $t<10^{-32}s$. So $d_{max}(t)$ can be preserved bigger than $D(t)$ from $10^{-32}s$ to $t_0$, including $t_{\gamma d}$. Then the horizon problem is safely removed. Since $\rho$ is approximately constant during inflation, $\epsilon(t)$ would decrease by $e^{100}$ times and flatness problem would disappear.


\section{A model of eternal accelerated expansion without particle horizon}
First, we present a brief review of our previous work \cite{8}.
Motivated by Padmanabhan \cite{9}, we first studied the de Sitter universe from emergence of space, and found the important character 
\begin{equation}
\label{7}
\rho +3p =-C,
\end{equation}
where $C$ is a positive constant. Then, we generalized this equation beyond the de Sitter universe. In this case, according to Eq.~\eqref{3} and continuity equation $\overset{.}{\rho}+3H(\rho+p)=0$, one can get the solutions of $\rho$ , $p$ , $H$ and $a$. Among the solutions we obtained, we found a model which can satisfy the initial condition $a(0)=0$ and is asymptotically de Sitter. It has following characteristics:
\begin{equation}
\label{q1}
\rho= \frac{B}{2}a^{-2}+\frac{C}{2}\ \ \ ,
\end{equation}

\begin{equation}
p=-\frac{B}{6}a^{-2}-\frac{C}{2}\ \ ,
\end{equation}

\begin{equation}
\label{q2}
H=\frac{2\alpha}{e^{2\alpha t}-1}+\alpha\ \ ,
\end{equation}

\begin{equation}
\label{q3}
a=A(e^{\alpha t}-\frac{1}{e^{\alpha t}}),
\end{equation}
where $\alpha=\sqrt{\frac{4\pi C }{3}}$ and $B$ is a positive integral constant (In our last paper, $B$ has been absorbed in $a$ for convenience. In this paper, we restore the $B$ for later discussion.). For detailed calculation, see \cite{8}. Since $C/2$ in our model (Eq.\eqref{q1}) is constant, it can be regarded as vacuun energy.  According to $\alpha=\sqrt{\frac{4\pi C }{3}}$,  Eq.\eqref{q2} can provide an evaluation of vacuum energy from $H$ and $t$. And based on the experimental data $H_0$ and $t_0$, we got $\alpha \approx 10^{-18} s^{-1}$ and $C/2 \approx 10^{-27}kg/m^3$, which is approximate to the experimental data.

According to Eq.~\eqref{q3} (with $\alpha \approx 10^{-18} s^{-1}$), it can be found
that the model is not capable of including an inflation for the early universe, which was first proposed to solve the horizon and flatness problems. However, as we will show, this model is actually free of horizon and flatness problems even without inflation at very early time, indicating that inflation may be unnecessary.

The horizon problem is caused by particle horizon. So, let us come back to Eq.~\eqref{4} and use Eq.~\eqref{q3}, we have 
\begin{equation}
\label{9} d_{max}(t)=a(t)\int_0 ^t \frac{dt'}{a(t')}=(e^{\alpha t}-\frac{1}{e^{\alpha t}})\int_0 ^t \frac{dt'}{(e^{\alpha t'}-\frac{1}{e^{\alpha t'}})}\ \ .
\end{equation}
The part of this definite integral is controlled by the indefinite integral  
\begin{equation}
\label{10} \int \frac{dt'}{(e^{\alpha t'}-\frac{1}{e^{\alpha t'}})}=\frac{1}{\alpha}\int \frac{dx'}{x'^2-1}=\frac{1}{2\alpha}\ln \left| \frac{x'-1}{x'+1} \right|
\end{equation}
that is divergent at $x'=1$($t'=0$). This means that $d_{max}(t)$ has no proper definition at any $t>0$, thus there is no particle horizon in this case, the horizon problem can be avoided. 

Since we have $\rho$ (Eq.~\eqref{q1}) and $a$ (Eq.~\eqref{q3}), according to Eq.~\eqref{2}, we may get the constraint on $k$ in our model. First, Eq.~\eqref{2} can be written as :
\begin{equation}
\label{w100} 
\rho=\frac{3H^2}{8\pi}+\frac{3k}{8 \pi a^2}=\frac{3\dot{a}^2+3k}{8\pi a^2}
\end{equation} 
Then, plugging Eq.~\eqref{q3} into this equation, one can get:
\begin{equation}
\label{q100} 
\rho=\frac{3\alpha ^2}{8\pi}+\frac{12A^2\alpha ^2 +3k}{8\pi a^2}= \frac{C}{2}+\frac{12A^2\alpha ^2 +3k}{8\pi} a^{-2}\ \ .
\end{equation}

So, when $k=0$ or $k=1$, one can get Eq.~\eqref{q1} directly. And for $k=-1$, to get Eq.~\eqref{q1}, we need $4A^2\alpha ^2 >1$.

As a first sight, there is no constraint on $k$ in our model, which seems to lead an existence of the flatness problem still. However, by plugging Eq.~\eqref{q1} into Eq.~\eqref{6}, one can get

\begin{equation}
\label{100} 
\epsilon(t)=\frac{3|k|}{8\pi \rho(t)a^2(t)}=\frac{3|k|}{8\pi \left(\frac{B}{2}a^{-2}+\frac{C}{2} \right)a^2(t)}=\frac{3|k|}{4\pi} \cdot \frac{1}{B+Ca^2(t)} \ \ .
\end{equation} 
The $\epsilon$ in our model will decrease as $a$ increases. This is quite different from the classic cosmological model Eq.~\eqref{s1}. So,
the present $\epsilon (t_0)$ can be close to 0 without supposing an initial value extremely close to zero, and  $\Omega_0$ would approximately be unity according to our definition of $\epsilon$. Thus, the flatness problem would disappear.


So far, we have showed that our model can solve the horizon and flatness problems. Since these problems can be solved by inflation theory, we will make a brief comparison between them. In most inflationary models, $ a(t)\propto e^{\chi t}$ and $\chi\approx 10^{34}$ s$^{-1}$. The expression of $a(t)$ (Eq.~\eqref{q3}) in our model also exists an exponential term $e^{\alpha t}$, and $\alpha \approx 10^{-18}$s$^{-1}$\cite{8}. So it does not lead to an early inflationary universe but a late de Sitter universe. And the way of solving the horizon and flatness problems is also different from inflationary model. 

\section{conclusions and discussions}
In our last paper, motivated by Padmanabhan, we proposed a cosmological model from emergence of space which is asymptotically de Sitter and can provide an evaluation of vacuum energy from the Hubble constant and the age of universe. However, there are two problems for this model, the absence of inflation at very early time and deceleration. As we know, the original inflation was proposed to solve the horizon and flatness problems. In this paper, we showed that our model is also free of horizon problem and flatness problem without inflation. We found that there is no particle horizon in our model, so the horizon problem can be avoided.
For flatness problem, our model gives a different result from the classic cosmological model. And the present universe can be flat without extremely initial condition in our model.

 While, there still remains one problem in our model, namely the nonexistence of deceleration phase. According to Eq.~\eqref{3}, $\rho +3p =-C$ means $\ddot a$ is always positive, and for expansion, it will cause a eternal accelerated expansion and our model is in this case. In most models of cosmology~\cite{2}, the universe is supposed to begin with a period of inflation, followed by a period of radiation dominance lasting until the time of radiation-matter equality, followed in turn by a period of matter dominance and then a period dominated by vacuum energy. The transition from deceleration (matter dominance) to acceleration has been found in the astronomical observation \cite{linder11}. To solve this problem, our model needs modification. And this will be our next step for the further research.

\section*{ACKNOWLEDGMENTS}

  We would like to express our deep gratitude to Wen-Yuan Ai and Hua Chen for enlightening discussion.  This work is supported by the National Natural Science Foundation of China~(Grant No.11171329).

  \bibliographystyle{apsrev4-1}
  \bibliography{reference}

\end{document}